\documentclass[journal,letterpaper]{IEEEtran_nssmic}
\usepackage{graphicx} 
\usepackage{subfigure}

\begin{document}

\renewcommand{\topfraction}{0.98}
\renewcommand{\bottomfraction}{0.98}
\renewcommand{\floatpagefraction}{0.5}

\title{Characterizing IMARAD CZT Detectors with Time Resolved Anode and Cathode Signals}
\author{Jeremy~Perkins*,~
        Henric~Krawczynski*,~
        and~Paul~Dowkontt*%
\thanks{*Department of Physics, Washington University, St. Louis, MO 63130}%
\thanks{Manuscript received October 29, 2003.}}
\maketitle

\begin{abstract}
We present the results of using standard IMARAD CZT detectors with a 100 MHz
readout of the anode and cathode pulses. The detectors,  2 cm x 2 cm
large and 0.5 cm thick, have 64 Indium pixellated anode contacts at a pitch
of 2.5 mm. We investigate the possibilities to improve on the
detector's photo-peak efficiency and
energy resolution using two depth of interaction (DOI) indicators:
(i) the total charge induced on the cathode and (ii) the drift time of
the electron cloud determined from the anode and/or the cathode pulses.
The DOI correction with the cathode charge gives better results,
increasing the 662 keV photo-peak efficiency by 57\% and improving the
energy resolution from 2.33\% to 2.15\% (FWHM, including the
electronic noise). The information
on the time dependence of the induced charge can be used as a
diagnostic tool to understand the performance of the
detector. Detailed comparison of the pulse
shapes with detector simulations gives the electron mobility and drift
time and makes it possible to assess the weighting potential and
electric field inside the detector.
\end{abstract}


\section{Introduction}

\PARstart{C}{admium} Zinc Telluride (CZT) has emerged as the detector material
of choice for X-rays and Gamma-rays because of its
large bandgap, excellent spatial and energy resolution, high stopping
power, and extremely high photo-effect cross section.  We are interested
in CZT detectors as focal plane detectors for future space-borne X-ray
and Gamma-ray telescopes, such as EXIST (Energetic X-Ray Imaging
Survey Telescope) \cite{Grindlay01} and ACT (Advanced Compton Telescope)
\cite{Kurfess}. In this paper we report results from testing standard
IMARAD detectors with a time resolved readout (Sections 2 and 3). We
investigated different options to correct the anode signals for the
depth of the primary interaction. A simplified detector
model and a comparison of simulated and measured signals are
presented in Section 4. A summary and outlook are given in Section 5.

\section{Equipment}
The data presented in these proceedings are for a standard Indium
contacted 64 pixel IMARAD CZT detector \cite{Imarad}. Previous
measurements of IMARAD detectors have been reported by Li et al. \cite{Li},
Hong et al. \cite{Grindlay02} and Narita et al. \cite{Grindlay03,Grindlay04}. The detector
dimensions are 2 cm x 2
cm x 0.5 cm with the pixel size being 1.6 mm square at a 2.5 mm
pitch. The electronics included 4 channels (3 anode pixels plus the
cathode) to read out the pulses with an analog bandwidth of 100
MHz similar to the setup described in \cite{Matteson01}. Using fast electronics enabled measurement of the drift time of
electrons in the detector to a resolution
of 10 ns.  In addition, we used a 16 channel ASIC \cite{eV}, which
allows us to measure pulse height information for 16 additional pixels. We plan to
use the ASIC readout to study events
occurring between pixels. The low noise design of the electronics used
batteries for power and batteries for the
high-voltage supply, cooled external fets, and ground isolation. The
100 MHz signal from the detector is digitized by a 500 MHz
oscilloscope and read into a PC for analysis.

\section{Measurements}

\begin{figure}
  \centering
  \includegraphics[width=2.5in]{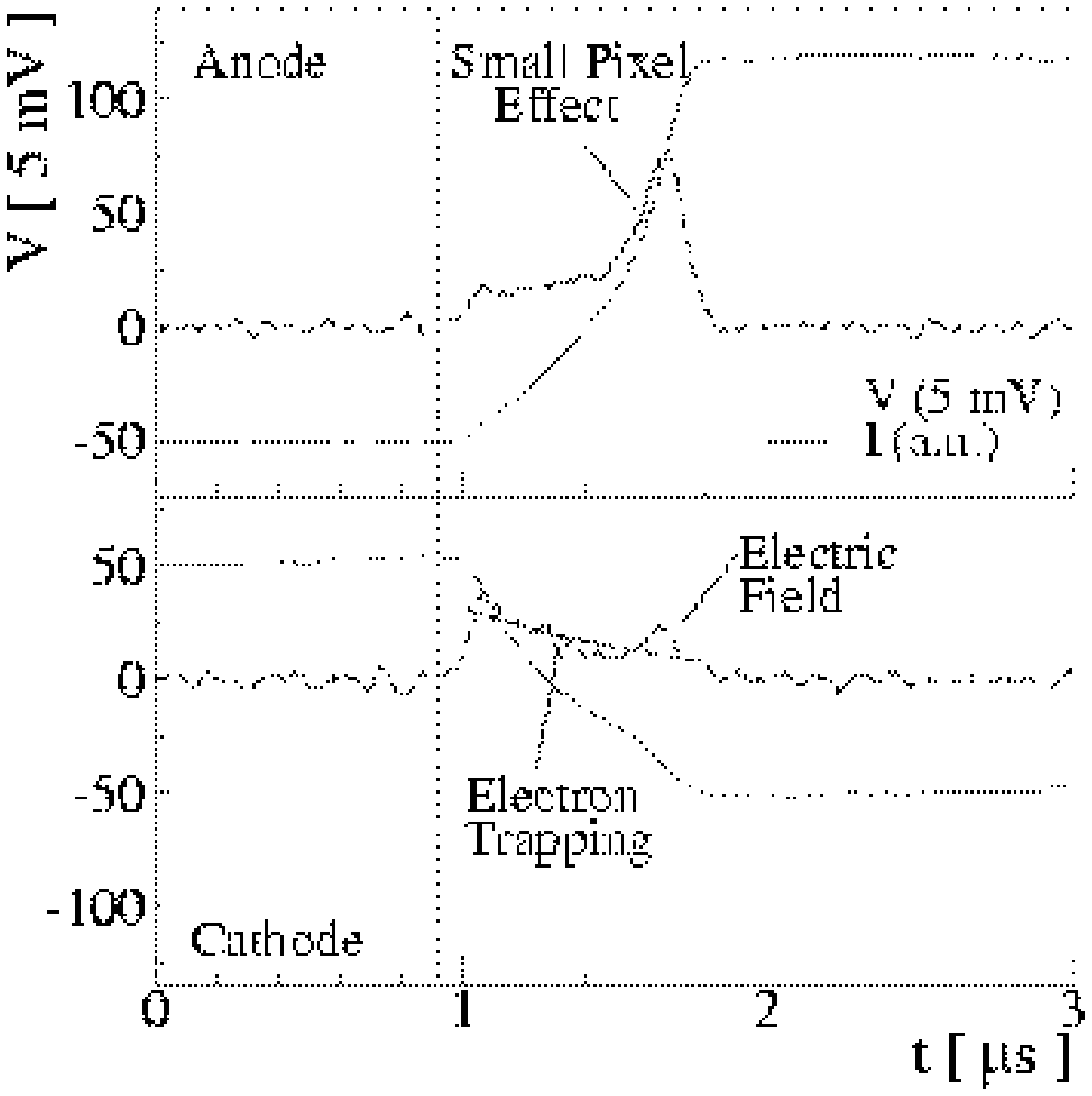}
  \caption{Pulse Shape Analysis.  In both plots, the solid line is
  proportional charge induced on the contact while the dashed line is the
  time derivative of that charge.  The top panel shows the
  signal as seen by an anode pixel and the bottom by the cathode.  The
  vertical lines mark the beginning and end of the pulse.  One can
  clearly recognize the small pixel effect, electron trapping, and electric
  field effects. The detector was biased at -500 V}
  \label{pulse}
\end{figure}

\subsection{Pulse Shape Analysis}
The following measurements were made using a Cs$^{137}$ source which emits an
X-ray line at 662 keV.
Figure \ref{pulse} shows a single pulse induced by a photon interacting with the
detector. The CZT is irradiated from the cathode side. While low-energy
photons do not penetrate far into the detector, most 662 keV photons
do not interact with the CZT; the few interactions are
basically homogeneously distributed within the detector volume.  The
detector is biased at -500 V on the cathode with each of the pixels
held at ground. The electrons drift at an almost constant
velocity. Shortly before impinging on the pixels, they induce most of
the charge. The small pixel effect (\cite{Barret,Luke}) is seen in Figure
1 by a peak in the anode current (dashed line) toward the
end of the pulse. The cathode does not show this effect. For many
pulses we observed an increase of the cathode current towards the end of
the pulse.  This increase may correspond to an increased electric
field near the anode pixels.  However, the effect is not yet fully
understood. The electron trapping is seen in an exponential decrease of
the current on the cathode.

\subsection{Cs$^{137}$ Spectra}
From data as shown in Figure \ref{pulse}, we determined the drift time of the
electron cloud in the detector, and the amplitudes of the anode and
cathode pulses.  The drift time is directly correlated with the depth of interaction in the
detector (\cite{Matteson, He}).  We used a Cs$^{137}$ source to produce several thousand
events in the detector and read them with a PC.

\begin{figure}
 \includegraphics[width=3in]{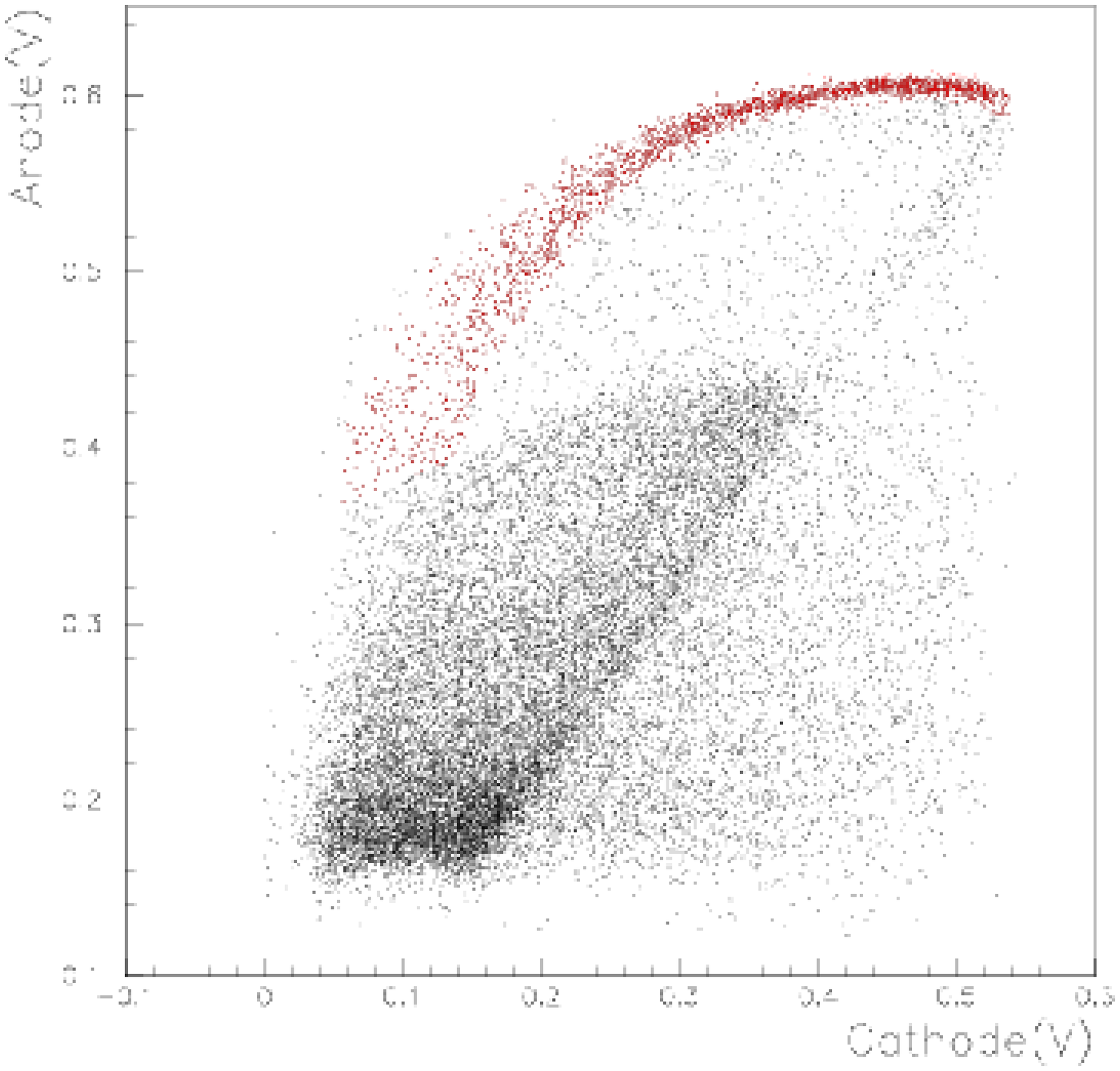}
  \caption{Cs$^{137}$ spectrum. Anode charge versus Cathode Charge.  The
  curved line is the 662 keV line.  Without correction, this curvature
  decreases the resolution of the line. The Compton continuum is also
  apparent.}
  \label{anode-cathode}
\end{figure}

\begin{figure}
  \centering
  \includegraphics[width=3in]{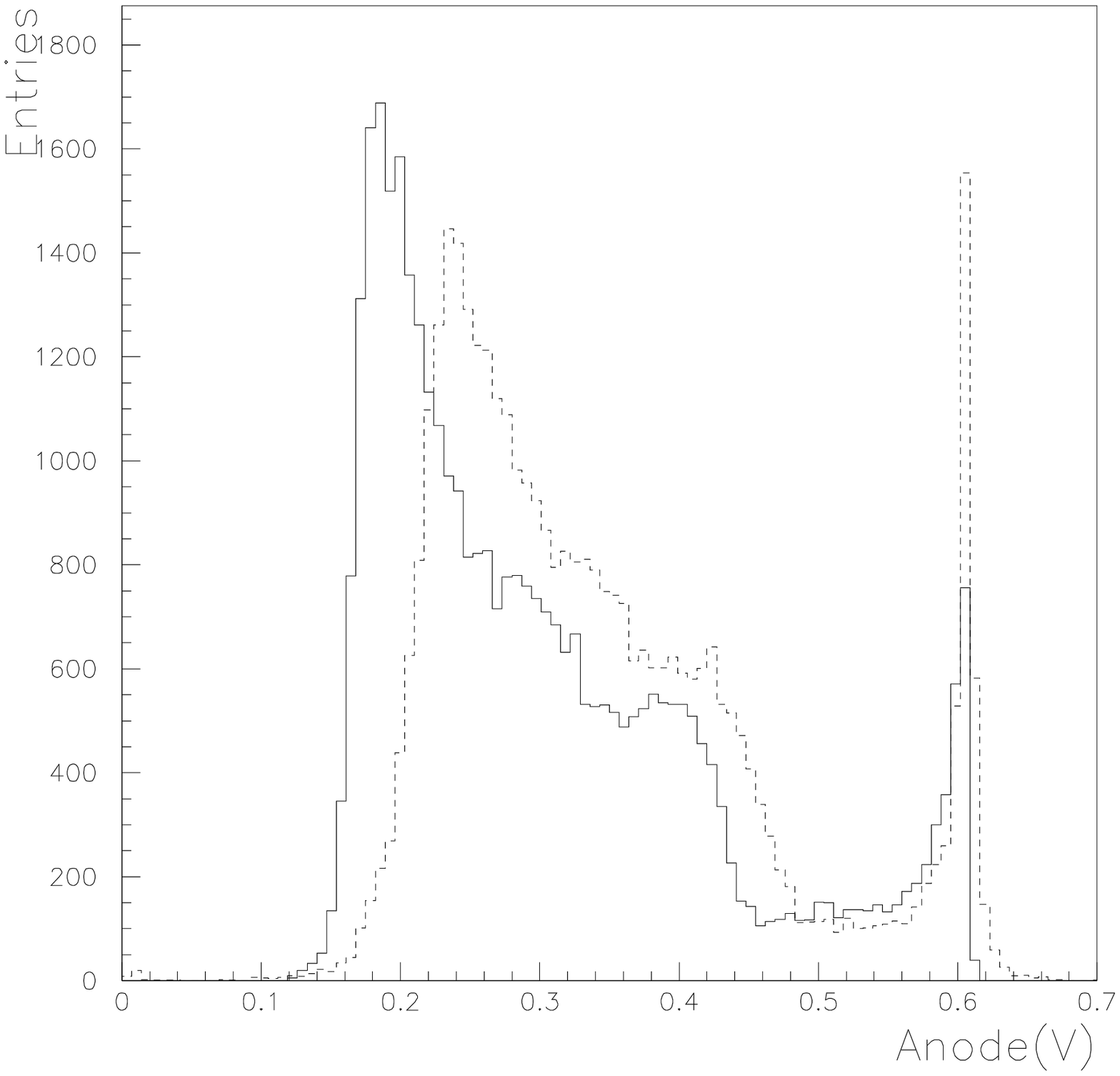}
  \caption{Cs$^{137}$ spectrum (662 keV).  The solid line is the uncorrected spectrum
  and the dashed line is the spectrum corrected for the charge
  curvature as seen in Figure \ref{anode-cathode}. This improves the FWHM from 2.33\% to
  2.15\%. The photopeak efficiency has also been increased 57\%.}
  \label{anode-cathode-hist}
\end{figure}

\begin{figure}
  \centering
  \includegraphics[width=3in]{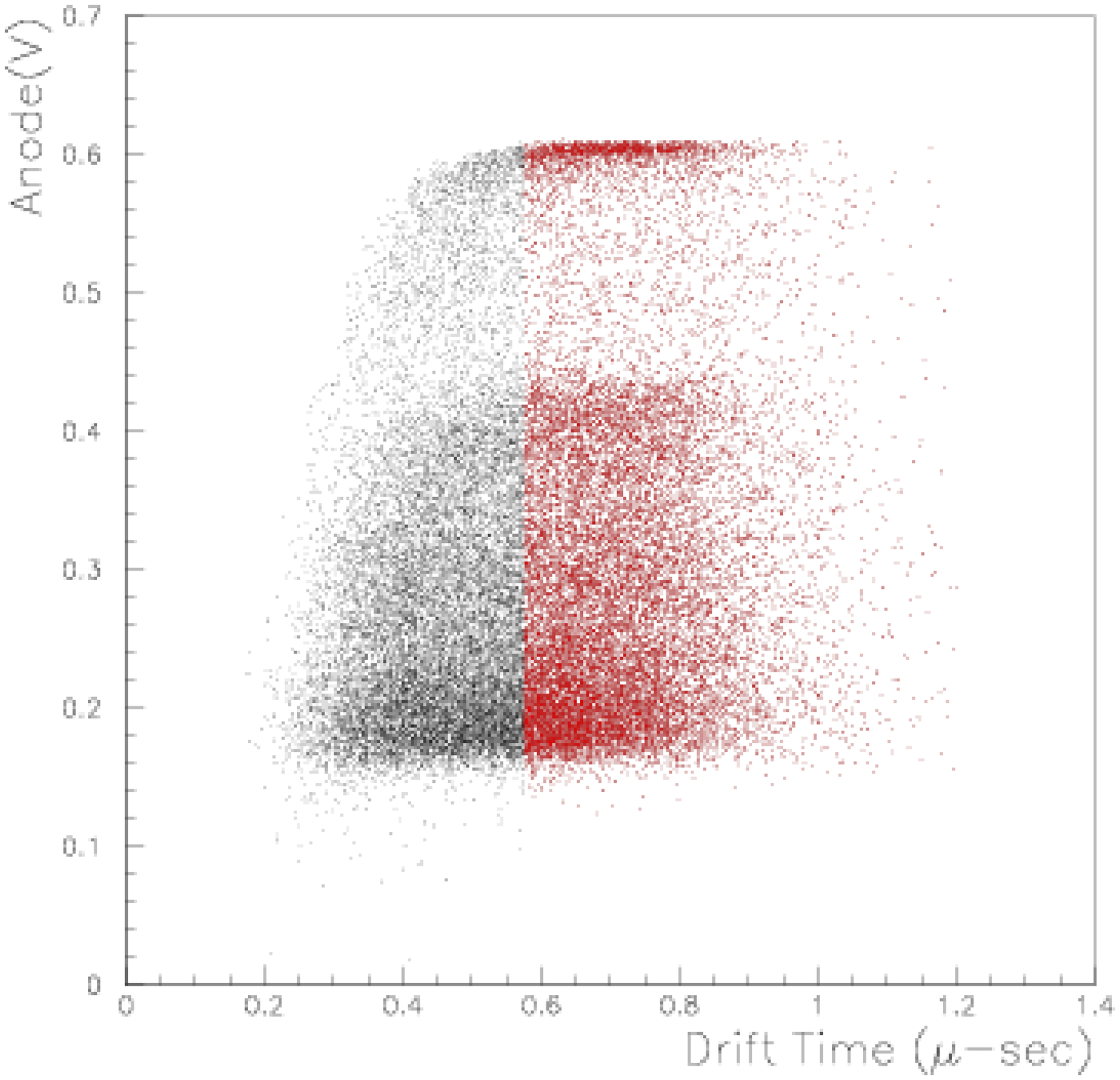}
  \caption{Cs$^{137}$ spectrum Anode charge versus Drift Time. The 662
  keV line can be recognized. The Compton continuum is also apparent. Notice
  that the photo-effect line widens for smaller drift
  times. In Figure \ref{anode-drift-hist} we have selected events with
  drift times greater than 0.57 $\mu$-sec shown in red in the plot above.}
  \label{anode-drift}
\end{figure}

\begin{figure}
  \centering
  \includegraphics[width=3in]{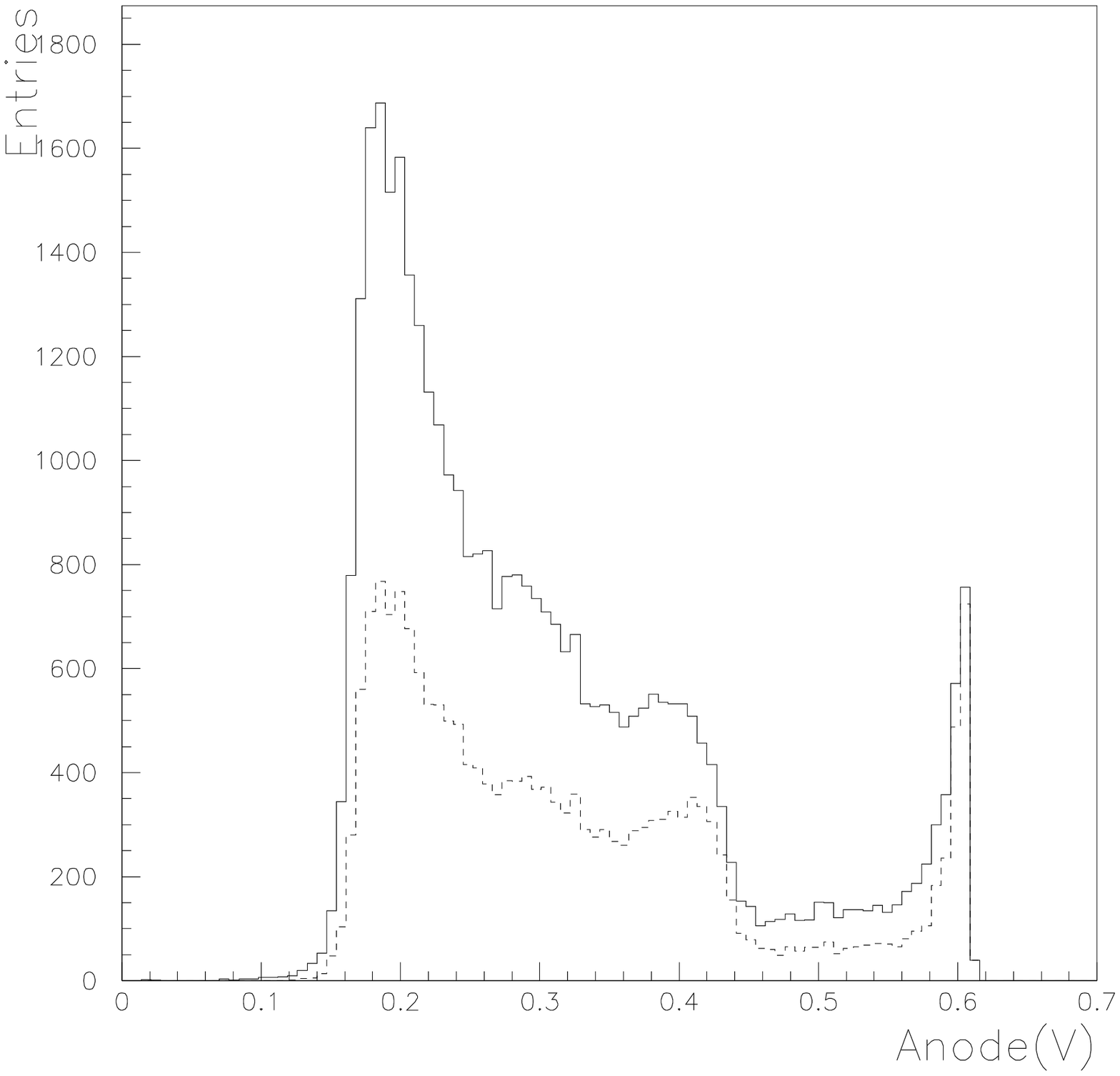}
  \caption{Cs$^{137}$ spectrum.  The solid line is the uncorrected spectrum
  and the dashed line is the spectrum after selecting events with
  drift times longer than 0.57 $\mu$-sec.}
  \label{anode-drift-hist}
\end{figure}

Figure \ref{anode-cathode} is a plot of the cathode charge versus the
charge induced on an anode pixel. 662 keV photo-effect events produce a
clearly recognizable arc in Figure \ref{anode-cathode}. A histogram of the anode signals
is shown by the solid line in Figure \ref{anode-cathode-hist}. The
photo-peak has a FWHM of 2.33\%. Correcting the
anode signal for the DOI with the cathode signal improves the energy
resolution to 2.15\% and increases the photopeak efficiency by
57\% (see the dashed line in Figure \ref{anode-cathode-hist}). Similar
results have been reported by Hong et al. \cite{Grindlay02}. By subtracting the electronic
resolution in quadrature we obtained a pre-correction energy resolution
of 1.8\% and a post-correction energy resolution of 1.6\%.  

Figure \ref{anode-drift} presents the same data showing the anode amplitude versus
the drift time. The photo-peak is obvious but
there is no clear correlation of amplitude and drift time which could
be used to correct the anode signals for DOI effects.  Selecting
events with drift times larger than 0.57 $\mu$-sec, improves the
energy resolution from 2.33\% to 2.12\% (1.5\% after subtracting electronic
noise) but the photo-peak efficiency decreases by 14\% (see Figure \ref{anode-drift-hist}).

\section{Simulations}
\begin{figure}
  \centering
  \includegraphics[width=3in]{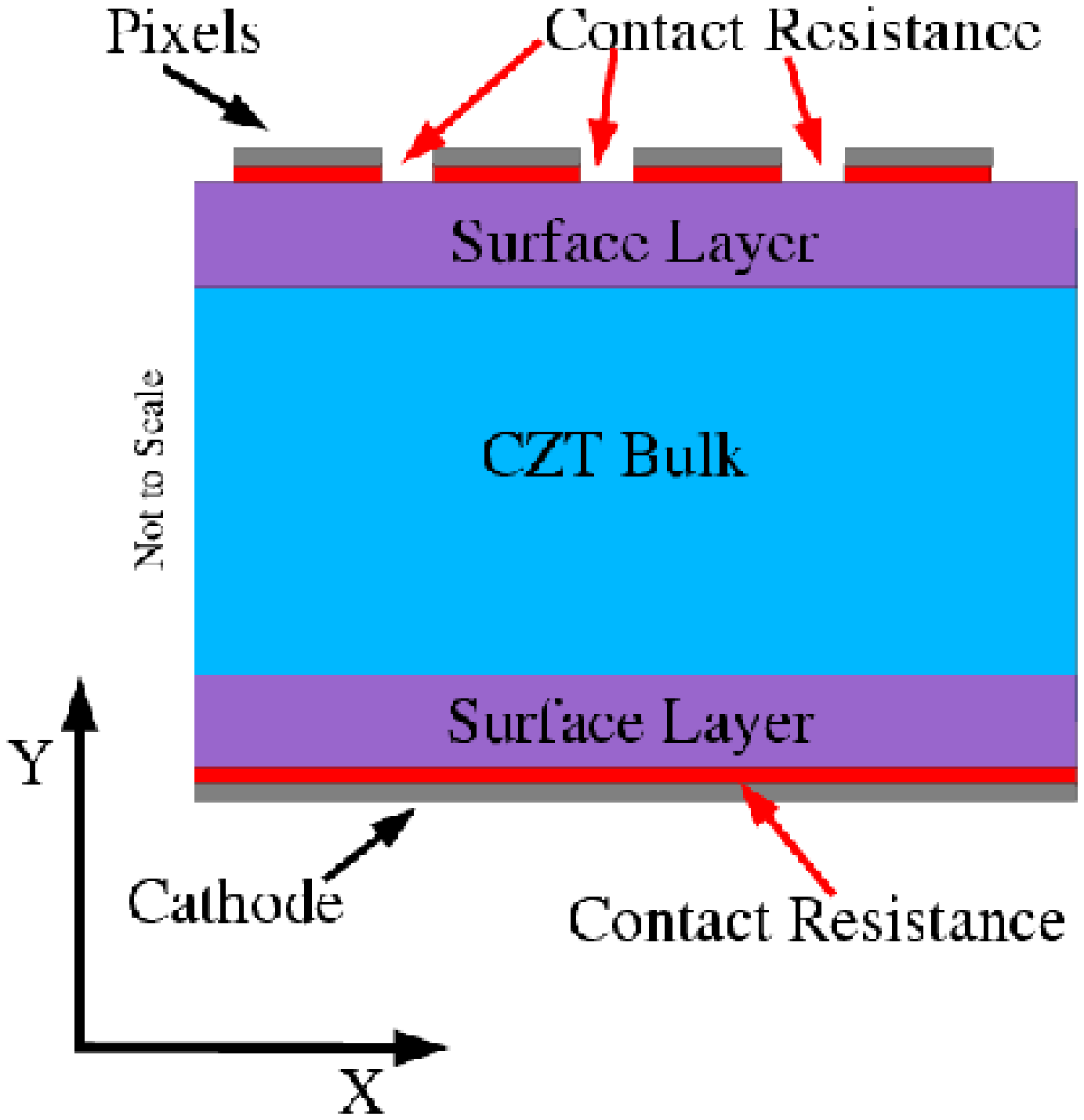}
  \caption{Graphical representation of the simulation model.  There
  are contact electrodes and a cathode on top of a layer that
  simulated contact resistance.  There are also two surface layers and
  a CZT bulk layer.}
  \label{model}
\end{figure}

The simulations are based on a two-dimensional model.  The pixel width corresponds
to that of the standard IMARAD detector. We chose a slightly larger
pixel pitch to mimic the 3-D geometry with our 2-D model. The weighting potential for each
pixel and the electric field inside the detector were determined using
a commercial semiconductor device simulator package called
ATLAS \cite{Atlas}. We modeled the CZT detector with a bulk region and
several layers to mimic surface conductivity and contact resistance.
The bulk is doped with $1.5\times10^6$ electrons/cm$^{3}$ We used surface layers with
a higher electron concentration than the bulk to simulate the surface
conductivity \cite{Bolotnikov01}. We also included significant contact resistance based on
pixel-cathode and pixel-pixel I-V measurements. The contact
resistivity of individual pixels is on the order of several
G$\Omega$. Figure \ref{model} shows a diagram of the model.

The ATLAS simulations give us the weighting potential for all contacts
and the electric field under realistic bias conditions. The weighting
potentials and electric field were used with our own code  to track
electrons though the detector and  to produce events similar to real
ones. The velocity of the electrons is determined by the electric field
and the assumed mobility.  

\begin{figure}
  \centering
  \includegraphics[width=3in]{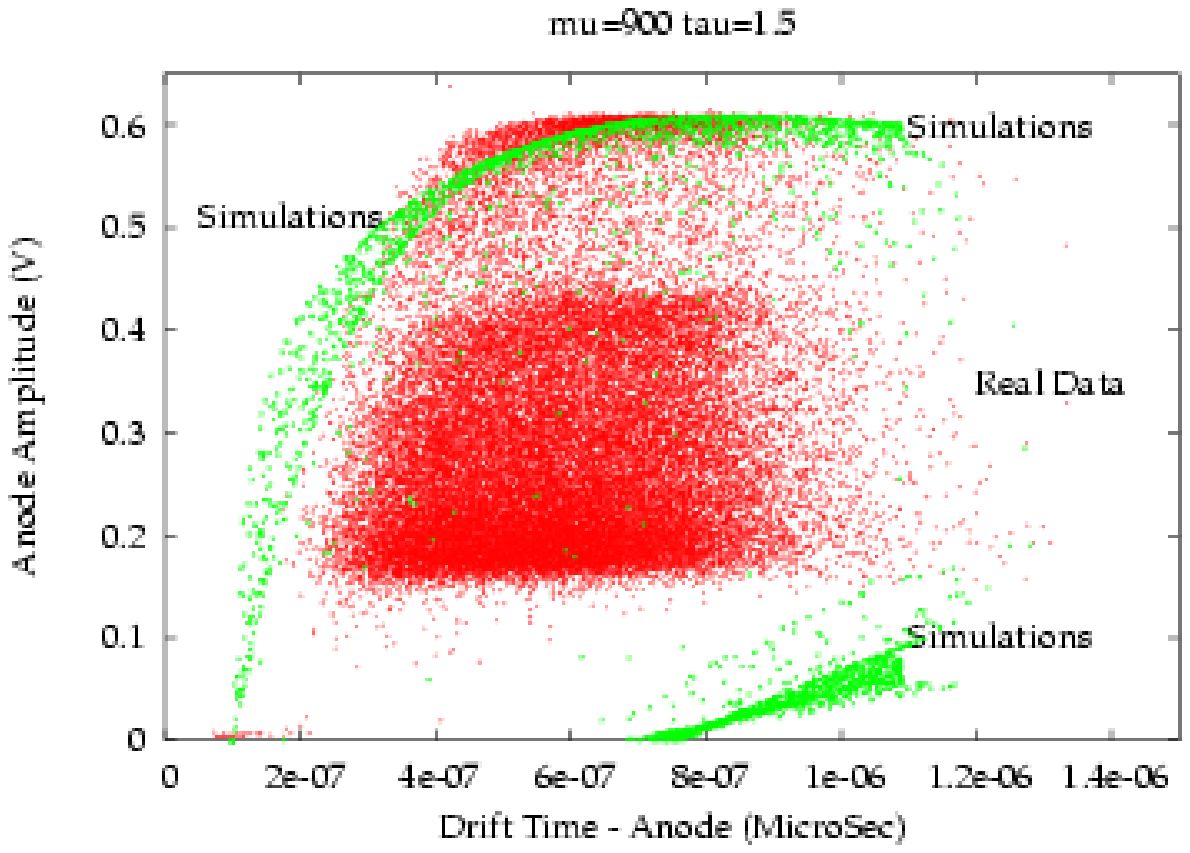}
  \caption{Comparison of Simulated and Experimental Data. Events at
  long-drift times are simulated well while events at short-drift
  times are not.  We suspect that these events are occurring
  between pixels. While there are only photo-effect events in the
  simulations, the real data shows also the Compton continuum.}
  \label{sims}
\end{figure}

Figure \ref{sims} compares a simulated spectrum to the actual data. The
mobility and trapping parameters of the simulations were chosen to
optimize the agreement between the simulations and the data.  The mobility is
given directly by the longest drift times while the trapping is
constrained by the slope of the photo-peak line (longer drift time
events will produce less charge).  The simulations deviate from the data in the
region of short drift times. We found
that these events primarily have interaction sites between
pixels (see Figure 8). The results indicate the electric field and/or the weighting
potential are not simulated correctly between pixels.  Other areas of
improvement might include the contact properties and the conductive
surface region.  Also, an extension from a two-dimensional model to a
three-dimensional one will no doubt improve the simulation. Based on
these simulations $\mu = 900~\rm cm^2V^{-1}s^{-1}$
and $\tau = 1.9~\mu$-sec. Previous studies of IMARAD detectors
have shown similar results \cite{Li}. The results depend strongly on
the assumed contact resistance.  Further 3 point measurements of the
latter are underway to improve our estimate.

\begin{figure}
  \centering
  \includegraphics[width=3in]{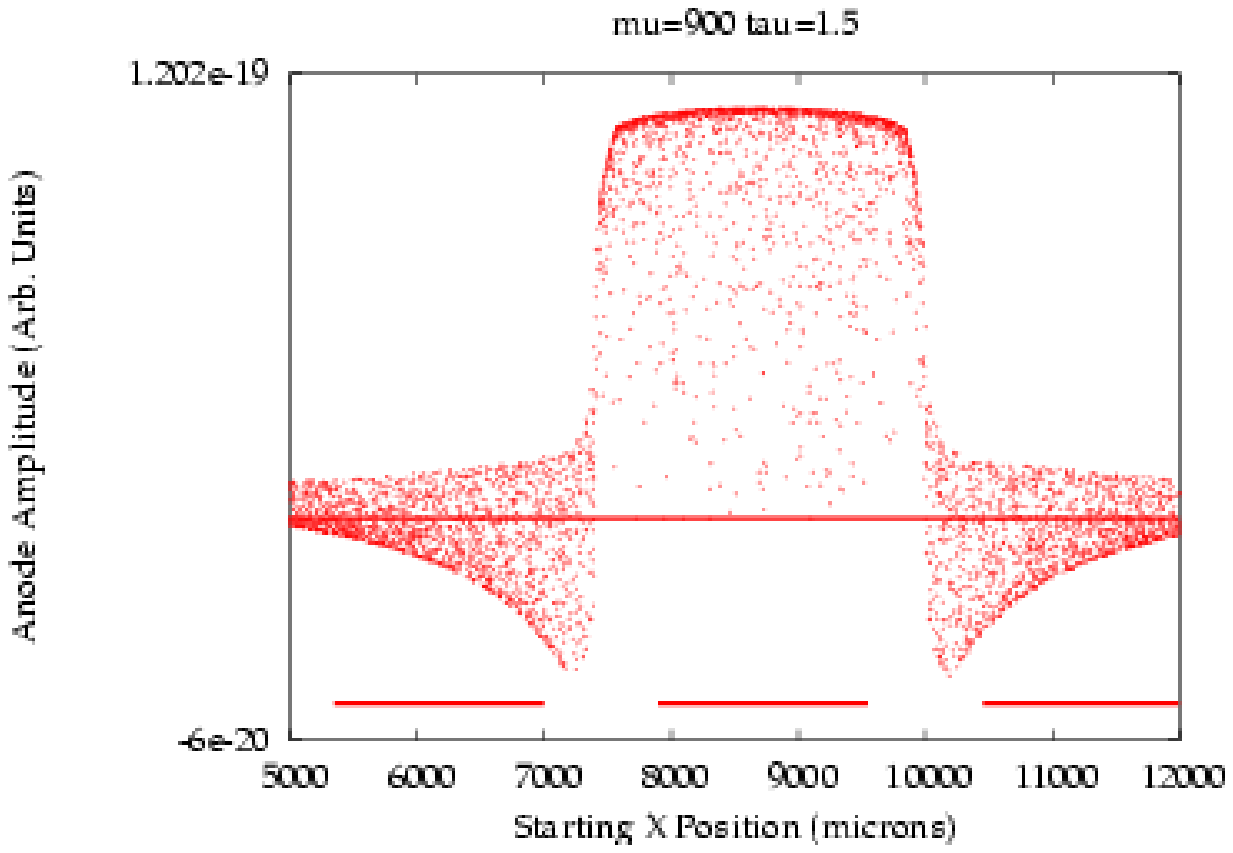}
  \caption{Charge Induced vs. Starting Position.  The horizontal lines
  at the bottom represent pixel locations.  The most charge is induced
  on the center pixel when events occur directly above the pixel while
  less or even negative charge is induced for off pixel events. See
  Figure 8 for a description of the orientation of the axis.}
  \label{xstart}
\end{figure}

Figure \ref{xstart} presents the charge induced on the anode versus the initial
interaction position.  In this figure, the pixels are shown as
horizontal lines along the bottom with the pixel under observation being the middle
line.  This only shows the horizontal interaction position and not
the depth of interaction (that can be seen in Figure \ref{sims} since there is
a direct relation between interaction depth and drift time).  The
events occurring directly above the pixel induce the most charge
while those events occurring off pixel induce less.  This indicates
there is charge sharing between neighboring pixels which has been
seen in other CZT detectors \cite{Bolotnikov02}.  There are also
events occurring above other pixels that induce negative charge on
the observed pixel. In the real data, events with low anode charges are
masked by the Compton events and are not observed. 

\section{Summary and Outlook}
In summary we have made measurements of the photo-peak of a 662 keV
Cs$^{137}$ line using standard Indium contacted IMARAD detectors giving a
FWHM of 2.15\% (1.6\% after subtracting the electronic contribution).
These detectors are substantially less expensive than standard
high-pressure Bridgman CZT and are thus extremely promising for experiments
requiring large detector areas. The correction
with the cathode amplitude produces better results than the correction with
drift times. Time resolved measurements of events allow for
development and fine tuning of simulations which subsequently allows
for optimization of new detectors.

Astrophysical applications require CZT detectors with broad energy
coverage.  However, at photon energies below ~100 keV leakage currents
deteriorate the performance of IMARAD detectors.  Reduced leakage
currents have been achieved with Au and Pt contacts
(\cite{Grindlay02,Grindlay03}). We next intend to improve on these results
by systematically testing high work function metals (Pt, Cr, Ni, etc.)
in combination with different surface preparation and passivation
methods.  The detectors are produced in a class-100 clean room.  We
use standard photolithographic methods to pattern the anode sides.
Results with novel contact and steering grid geometries will be
presented in a forthcoming paper.

\section*{Acknowledgment}
The authors would like to thank Uri El Hanany of IMARAD for supplying
us with crystals, Jim Matteson of UCSD for discussing the electronics, and our
shop technicians for manufacturing the equipment.

\end{document}